\documentclass{Interspeech}
\usepackage{utfsym}
\usepackage{multirow}
\usepackage{cite}
\usepackage{amsmath,amssymb,amsfonts}
\usepackage{algorithmic}
\usepackage{graphicx}
\usepackage{textcomp}
\usepackage{booktabs}
\usepackage{ulem}
\usepackage{array}
\usepackage{amsmath}  

\newcommand{\Xhline}[1]{\noalign{\hrule height #1}}

\usepackage{tabularx} 
\newcolumntype{Y}{>{\centering\arraybackslash}X} 

\interspeechcameraready

\title{Towards LLM-Empowered Fine-Grained Speech Descriptors\\ for Explainable Emotion Recognition}

\author[affiliation={1}]{Youjun}{Chen}
\author[affiliation={2*}]{Xurong}{Xie}
\author[affiliation={1}]{Haoning}{Xu}
\author[affiliation={3}]{Mengzhe}{Geng}
\author[affiliation={1}]{Guinan}{Li}
\author[affiliation={1}]{\\Chengxi}{Deng}
\author[affiliation={1}]{Huimeng}{Wang}
\author[affiliation={1}]{Shujie}{Hu}
\author[affiliation={1*}]{Xunying}{Liu}

\affiliation{}{The Chinese University of Hong Kong}{Hong Kong SAR, China}
\affiliation{Institute of Software}{Chinese Academy of Sciences}{China}
\affiliation{}{National Research Council Canada}{Canada}
\email{\{yjchen, xyliu\}@se.cuhk.edu.hk, xurong@iscas.ac.cn}
\keywords{Speech emotion recognition, Speech emotion descriptors, LLM, VAE, Information Bottleneck}

\usepackage{comment}
\begin{document}

\maketitle
\renewcommand{\thefootnote}{*}%
\footnotetext{Corresponding author.}%
\renewcommand{\thefootnote}{\arabic{footnote}}%

\begin{abstract}
This paper presents a novel end-to-end LLM-empowered explainable speech emotion recognition (SER) approach. Fine-grained speech emotion descriptor (SED) features, e.g., pitch, tone and emphasis, are disentangled from HuBERT SSL representations via alternating LLM fine-tuning to joint SER-SED prediction and ASR tasks. VAE compressed HuBERT features derived via Information Bottleneck (IB) are used to adjust feature granularity. Experiments on the IEMOCAP and MELD benchmarks demonstrate that our approach consistently outperforms comparable LLaMA-based SER baselines, including those using either (a) alternating multi-task fine-tuning alone or (b) feature disentanglement only. Statistically significant increase of SER unweighted accuracy by up to 4.0\% and 3.7\% absolute (5.4\% and 6.6\% relative) are obtained. More importantly, emotion descriptors offer further explainability for SER.
\end{abstract}


\vspace{-0.2cm}
\section{Introduction}

\vspace{-0.1cm}
Speech emotion recognition (SER) is fundamental to human-computer interaction.
Despite decades of promising advancements, most research \cite{shen2024emotion,naini2024generalization,Tzeng2025,conf/icassp/DangVNW23} has primarily focused on classifying speech into single, discrete emotion categories.
However, such approaches oversimplify the complexity of human emotions, as single-label outputs fail to capture the nuanced interplay of linguistic content and paralinguistic cues. Consequently, traditional SER systems lack the granularity and explainability necessary for a deeper and more human-like understanding of emotional expression~\cite{xu2024secap}.

In recent years, substantial efforts have been directed toward improving the explainability of the SER models by probing linguistic content- and emotion descriptor-related intermediate representations and separately decoding them.
These include:
\textbf{1)} disentangling task-oriented intermediate features from compact encodings\cite{yuan2024,mohebbi2024disentangling,si2021variational,wu2024multi,ma2024leveraging,wang2021variational}, e.g. SSL representations; and \textbf{2)} modeling them individually based on the input audios and transcripts generated by ASR models \cite{schuller2004speech,gao2024speech,hsu2023,makiuchi2021multimodal,xi2022frontend,jin2015speech}.
More recent research has focused on large language models (LLMs), demonstrating their ability to effectively interpret both linguistic content and emotion descriptors in SER tasks even as many approaches continue to rely on single-word labels for entire utterances \cite{tang2023salmonn,chu2023qwen,hu2024wavllm,kyung2024enhancing,yang2024large}. Furthermore, LLMs have been employed to generate natural language descriptions, providing more precise and accessible insights into fine-grained emotion descriptors \cite{mei2024wavcaps,deshmukh2024training,xu2024secap,liang2024aligncap,lin2024clap4emo}.

However, these prior studies suffer from the following limitations: 
\textbf{a)} Lack of explainable SER methods using both feature disentanglement and fine-grained emotion descriptors. While some methods disentangle content- and emotion descriptor-related representations \cite{yuan2024,mohebbi2024disentangling,si2021variational,wu2024multi,ma2024leveraging,wang2021variational,schuller2004speech,gao2024speech,hsu2023,makiuchi2021multimodal,xi2022frontend,jin2015speech} or associate emotion-related emotion descriptors with predicted emotion labels \cite{mei2024wavcaps,deshmukh2024training,xu2024secap,liang2024aligncap,lin2024clap4emo}, none integrates these approaches to improve human understanding of emotion cues; 
\textbf{b)} Lack of an end-to-end explainable SER architecture capable of simultaneously generating transcript, emotion descriptors and emotion labels directly from input speech. Instead, some methods rely on ASR models to transcribe input speech into text before performing SER \cite{schuller2004speech,gao2024speech,hsu2023,makiuchi2021multimodal,xi2022frontend,jin2015speech}, making the SER performance inherently dependent on the efficacy of the selected ASR models;
and \textbf{c)} Lack of exploration of the tight connection between fine-grained emotion descriptors prediction and SER performance. While some recent studies have focused on capturing fine-grained emotion descriptors, their quantitative impact on SER performance remains underexplored \cite{mei2024wavcaps,deshmukh2024training,xu2024secap,liang2024aligncap,lin2024clap4emo}.

To this end, this paper presents a novel end-to-end LLM-empowered explainable SER approach. Fine-grained emotion descriptor features are disentangled from HuBERT SSL representations via alternating LLM fine-tuning to joint SER-SED prediction and content transcription based ASR tasks. VAE compressed HuBERT features derived via IB are used to adjust the trade-off between accuracy and model complexity. 

\begin{figure*}[htbp]
\centering
\setlength{\abovecaptionskip}{0pt plus 1pt minus 3pt}
\includegraphics[scale=0.41]{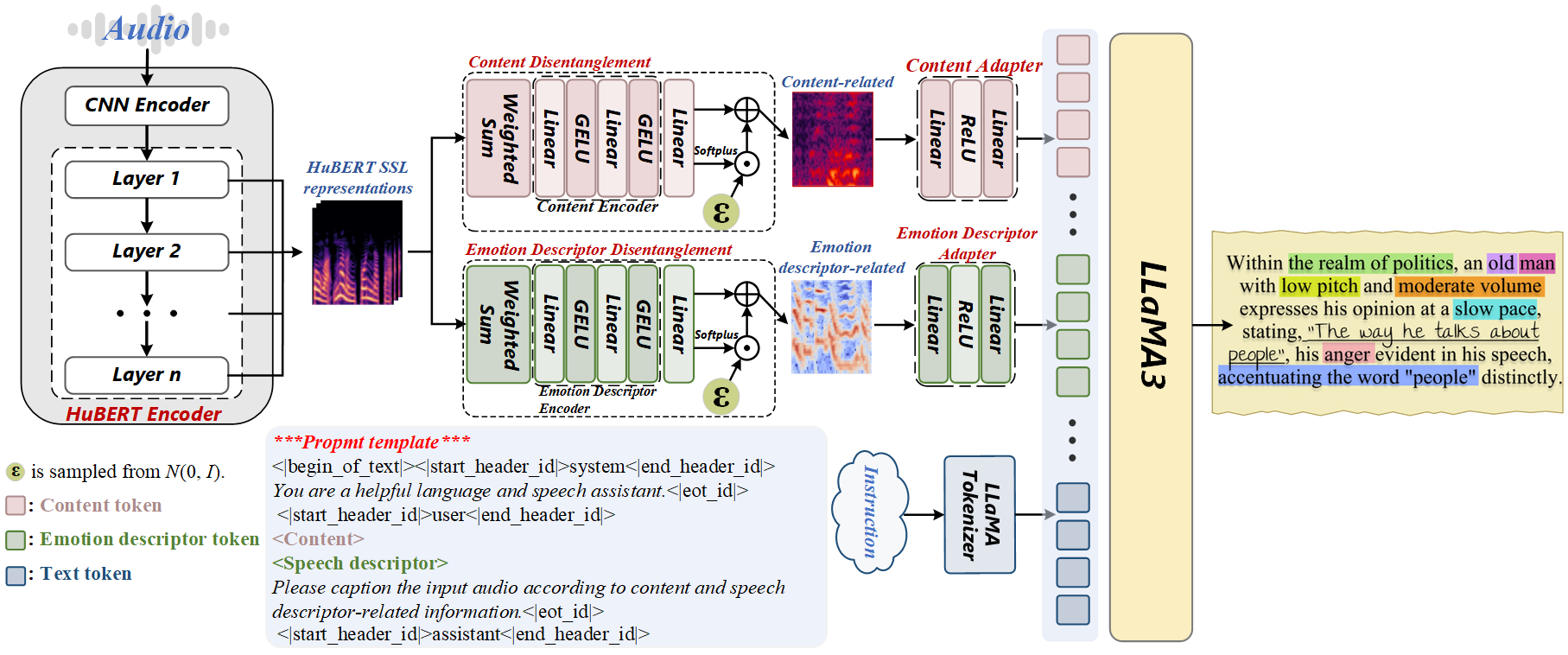}
\caption{Illustration of the proposed LLM-empowered explainable speech emotion recognition approach and prompt template.}
\label{fig1}
\vspace{-0.5cm}
\end{figure*}

Experiments on the benchmark IEMOCAP and MELD datasets suggest our approach consistently outperforms the comparable LLaMA SER baselines constructed using either a) alternating multi-task fine-tuning alone; or b) features disentanglement only. Statistically significant increases of SER unweighted accuracy by up to 4.0\% and 3.7\% absolute (5.4\% and 6.6\% relative) are obtained. More importantly, the predicted emotion descriptors offer further explainability for SER. 

The main contributions of our work are summarized below:

\textbf{1)} To the best of our knowledge, this paper presents the first work to use both feature disentanglement and fine-grained speech emotion descriptors for explainable SER. In contrast, previous research paid attention to either feature disentanglement \cite{yuan2024,mohebbi2024disentangling,si2021variational,wu2024multi,ma2024leveraging,wang2021variational,schuller2004speech,gao2024speech,hsu2023,makiuchi2021multimodal,xi2022frontend,jin2015speech} or fine-grained emotion descriptor prediction \cite{mei2024wavcaps,deshmukh2024training,xu2024secap,liang2024aligncap,lin2024clap4emo} alone.

\textbf{2)} This paper introduces a novel end-to-end LLM-based explainable SER architecture that simultaneously generates transcript, emotion descriptors and emotion labels directly from input speech. In contrast, prior research employs external ASR models to generate transcripts before performing SER \cite{schuller2004speech,gao2024speech,hsu2023,makiuchi2021multimodal,xi2022frontend,jin2015speech}, which may lead to low SER performance affected by ASR errors.

\textbf{3)} This paper investigates the tight connection between fine-grained SED prediction and SER performance. In contrast, prior research focused on generating fine-grained and precise emotion descriptors alone without further quantitative research on the importance of fine-grained SED prediction for SER performance \cite{mei2024wavcaps,deshmukh2024training,xu2024secap,liang2024aligncap,lin2024clap4emo}. Our code is available at https://github.com/SEUJames23/explainable-emotion-recognition.

\vspace{-0.3cm}
\section{Feature Disentanglement}
\vspace{-0.1cm}
\subsection{Alternating Multi-task Fine-tuning}
\vspace{-0.2cm}
A parameter-efficient fine-tuning method known as low-rank adaptation, or LoRA \cite{hu2021lora}, is widely employed to fine-tune the LLM-based models. However, extensively fine-tuning the model using the LoRA method on a large amount of prompt-repetitive speech-text data may cause the model to overfit on specific speech tasks and fail to address the remaining instructions \cite{hu2024wavllm}. To avoid this problem, we adopt an alternating multi-task fine-tuning strategy during training. More details on leveraging alternating multi-task fine-tuning to disentangle content- and emotion descriptor-related representations can be found in Sec. \ref{twostages}.
\vspace{-0.3cm}
\subsection{Information Bottleneck}
\vspace{-0.1cm}
IB \cite{tishby2000information} aims to learn a compressed representation $\mathbf{Z}$ of the input $\mathbf{X}$ while preserving relevant information about the output $\mathbf{Y}$. The whole process can be interpreted as maximizing the mutual information $I\mathbf{(Z,Y)}$ between $\mathbf{Y}$ and $\mathbf{Z}$, and minimizing the mutual information $I\mathbf{(Z,X)}$ between $\mathbf{X}$ and $\mathbf{Z}$ simultaneously. The mathematical formula of IB loss is shown in Eq. (\ref{ib1}).
\vspace{-0.1cm}
\begin{align} 
\mathcal{L}_{IB} = \beta\cdot I(\mathbf{Z,X}) - I(\mathbf{Z,Y})
\label{ib1}
\end{align}

However, the exact computation of $\mathcal{L}_{IB}$ is intractable. To address this, a variational approximate estimate of IB (VIB) \cite{alemi2016deep} has been introduced. Specifically, for each training example $(\mathbf{x}, \mathbf{y})$, the IB loss is upper bounded by:
\vspace{-0.1cm}
\begin{align} 
\mathcal{L}_{VIB}=\mathbb{E}_{\mathbf{x}\sim p\mathbf{(x)}}\big[\mathbb{E}_{\mathbf{z} \sim p\mathbf{(z|x)}}[-\log q\mathbf{(y|z)}] & \notag \\
+ \beta \cdot \mathbb{KL}(p\mathbf{(z|x)}, p_0\mathbf{(z)}) \big]
\label{ib2}
\end{align}

where, $q\mathbf{(y|z)}$ is a variational approximation to $p\mathbf{(y|z)}$. $p_0\mathbf{(z)}$, a specified prior distribution for latent representation $\mathbf{Z}$, usually takes standard normal distribution. And $p\mathbf{(z|x)}$ denotes an estimate of the posterior probability of $\mathbf{z}$. $\beta \ge0$ controls the balance between compression and prediction.

In the VIB loss function, the first term, the task loss, encourages the encoder to preserve information relevant to the label, while the second term, the KL divergence, pushes it to discard as much information as possible. Eq.(\ref{ib2}) takes the same form of variational auto-encoder (VAE) style Bayesian learning \cite{kingma2013auto,xie2021bayesian,hu2021bayesian}, with the reparameterization trick used in the optimization process.
\vspace{-0.3cm}

\section{System Architecture}
\vspace{-0.1cm}
As is illustrated in Fig. \ref{fig1}, the architecture of our approach consists of a HuBERT encoder, two feature disentanglement blocks, two feature adapters and an LLM decoder. Here, we denote the HuBERT representations, intermediate representations, and target outputs as $\mathbf{X}$, $\mathbf{Z}$ ($\mathbf{Z^{Con}}$ for content-related, $\mathbf{Z^{Des}}$ for emotion descriptor-related), and $\mathbf{Y}$, respectively.
\vspace{-0.2cm}
\subsection{HuBERT Encoder}
\vspace{-0.2cm}
We use the encoder of the self-supervised learning (SSL) model HuBERT-large\footnote{https://huggingface.co/facebook/hubert-large-ls960-ft} \cite{journals/corr/abs-2106-07447} to encode the input audio. Since the different hidden layer outputs of the SSL models contain the latent representations for various tasks \cite{journals/corr/abs-2106-07447,9747490}, we use all the hidden layers' outputs as $\mathbf{X}$. We keep the HuBERT encoder's parameters frozen throughout the entire training process.
\vspace{-0.2cm}
\subsection{Content and Descriptor Features Disentanglement}
\vspace{-0.2cm}
Based on IB, we construct content and emotion descriptor disentanglement blocks consisting of weighted sum layers, content and emotion descriptor encoders, and VAE to probe the compact intermediate representations. Specifically, during training, weighted sum layers first learn the weights to extract the most relevant information from the SSL representations $\mathbf{x}$ based on the downstream tasks. After being passed through the corresponding encoders which comprise a two-layer perceptron with \rm{GELU} activation between the linear layers, $x$ are then input to two linear layers to yield the mean and standard deviation vectors, $i.e.$, $\mathbf{\mu(x)}$ and $\mathbf{\sigma(x)}$, of the posterior distribution $p(\mathbf{z}|\mathbf{x})$. We simulate Gaussian samples of $\mathbf{z}$  which are then fed into the LLM decoder. During inference, $\mathbf{z}=\mathbf{\mu(x)}$ is used instead of sampling from $p\mathbf{(z|x)}$.
\vspace{-0.2cm}
\subsection{Content and Descriptor Feature Adapters}
\vspace{-0.2cm}
To enable the LLM decoder to comprehend the content- and emotion descriptor-related information, we incorporate trainable adapters mapping the latent representations into the embedding space of the LLM. Following \cite{fang2024llama}, after downsampling $\mathbf{Z}$, the corresponding adapters produce the textual content and emotion descriptor embeddings $\mathbf{S}$ ($\mathbf{S^{Con}}$ for content and $\mathbf{S^{Des}}$ for emotion descriptor) with a two-layer perceptron with ReLU activation between the linear layers. The above process can be formalized as Eq. (\ref{eq3}).
\begin{align} 
\mathbf{S} = {\rm Linear} ({\rm ReLU}({\rm Linear}({\rm DownSample}(\mathbf{Z}))))
\label{eq3}
\end{align}
\vspace{-0.6cm}
\subsection{Large Language Model}
\vspace{-0.2cm}
We use the state-of-the-art Llama-3.1-8B-Instruct3\footnote{https://huggingface.co/meta-llama/Llama-3.1-8B-Instruct} \cite{dubey2024llama} as the LLM decoder, working as $\rm{log}\mathbf{q(y|z)}$ in VIB. The prompt template $\rm{Prompt}(\cdot)$ is shown in Fig. \ref{fig1} (bottom middle, in light blue), where $\mathbf{S^{Con}}$ and $\mathbf{S^{Des}}$ are filled into the position corresponding to $\langle Content\rangle$ and $\langle Speech\ descriptor\rangle$, respectively. Then the entire sequence $\rm{Prompt}(\mathbf{S})$ is input into the LLM decoder to autoregressively generate the text response $\mathbf{Y}^T=[\mathbf{y}_1^T, \mathbf{y}_2^T,\dots,\mathbf{y}_M^T]$. The LLM loss $\mathcal{L}_{LLM}$ of the decoder based on the cross-entropy loss can be regarded as the task loss in VIB loss (Eq. \ref{ib2}), which is calculated by: 
\begin{align} 
\mathcal{L}_{LLM} = -\sum_{i=1}^M \log p(\mathbf{y}_i^T| {\rm Prompt}(\mathbf{S}),\mathbf{Y}_{<i}^T)
\label{eq4}
\end{align}
\vspace{-0.5cm}

\subsection{Alternating Fine-tuning on ASR and SER-SED Tasks}\label{twostages}
\vspace{-0.2cm}
As shown in Fig. \ref{fig2}, we split the whole training process into two stages on the two sequential downstream tasks. In the first stage, the content encoder and adapter are trainable, while the emotion descriptor encoder and adapter are frozen with emotion descriptor tokens set to zero. Via LLM fine-tuning to speech content transcription based ASR task, content-related information disentangled from HuBERT SSL representations is preserved as much as possible. Following the first stage, the content encoder and adapter are frozen, while the emotion descriptor encoder and adapter are trainable in the second stage. Since sufficient content-related information has been retained in the first stage, the emotion descriptor encoder only focuses on disentangling SED-related information as a complementary in the joint SER-SED prediction and speech content transcription based ASR tasks. Note that we do not need to follow a two-stage process to obtain the content and emotion descriptor prediction during inference.

\vspace{-0.3cm}
\section{Experiment}
\vspace{-0.1cm}
\subsection{Datasets}
\vspace{-0.2cm}
We use publicly available SED prediction and SER datasets in the training and evaluation stages.

\noindent{\textbf{Training}. SpeechCraft \cite{jin2024speechcraft} annotates the utterances from four well-known corpora in a unified and fine-grained manner, which mainly considers the emotion descriptors such as pitch, energy, speed, age, gender, tone and emphasis. We select the large-scale subset GigaSpeech-m \cite{chen2021gigaspeech} among them as the training dataset, which includes above 670k audio clips with a total duration reaching 739.91 hours. Additionally, utterances regarded as happy, sad, angry and neutral accounted for the vast majority of these data, with a few surprised, disgusted and fearful ones. 

\noindent{\textbf{Evaluation}. Following \cite{liang2024aligncap}, we conduct zero-shot evaluations on the SED prediction performance of the proposed method on the Mandarin and English speech emotion caption datasets EMOSEC \cite{thuseethan2022emosec}, which consists of 15 male and 15 female speakers and 45\,039 utterances in total. Furthermore, we explore the SER performance of the proposed method on the two widely-used datasets, $i.e.$, IEMOCAP \cite{Busso2008} and MELD \cite{poria2018meld}. To be consistent and comparable with previous works \cite{ma2024emobox,yuan2024disentanglement} on IEMOCAP, we merge ``excited'' with ``happy'' to better balance the size of each emotion class, resulting in four classes. And we conduct leave-one-session-out 5-fold cross-validation on this data\footnote{More details on the data pre-processing can be found in \cite{ma2024emobox}.}. As for MELD, we just follow its original split setup.

\begin{figure}[htbp]
\centering
\setlength{\abovecaptionskip}{0pt plus 1pt minus 3pt}
\includegraphics[scale=0.26]{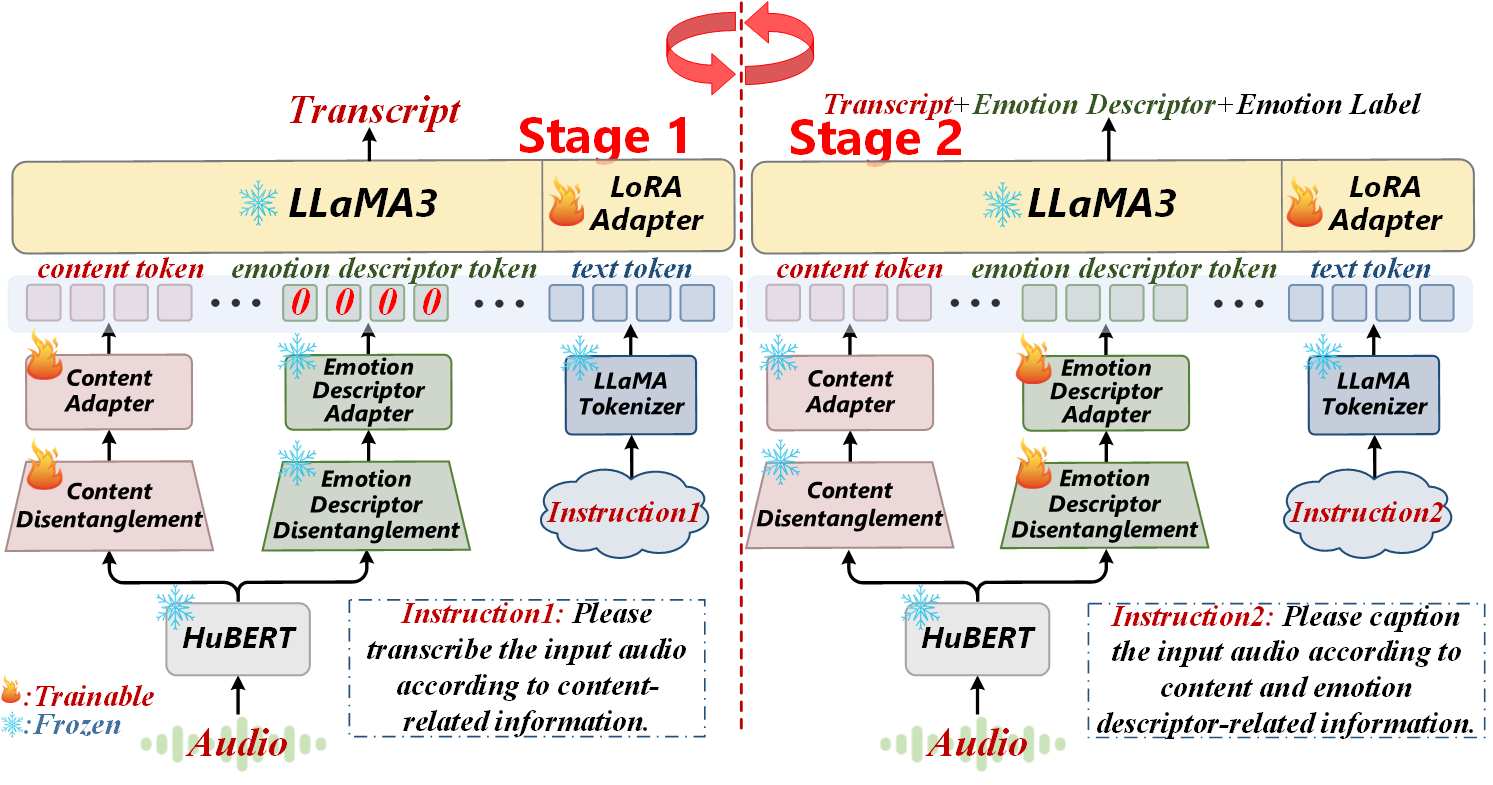}
\caption{Illustration of two stages of the whole training process on the two sequential downstream tasks with the alternating multi-task fine-tuning strategy.}
\label{fig2}
\vspace{-0.5cm}
\end{figure}

\vspace{-0.3cm}
\subsection{Baseline Systems}
\vspace{-0.2cm}
We compare the proposed method with the following baselines:

\noindent{\textbf{HuBERT+LLM-SER+SED+ASR}}: LLaMA SER system fine-tuned to joint SER-SED prediction and ASR tasks using standard HuBERT features.

\noindent{\textbf{HuBERT+Feature Disentanglement(FD)+LLM-SER+ASR}}:  LLaMA SER system fine-tuned to joint SER and ASR tasks using disentangled HuBERT representations. 

\noindent{\textbf{HuBERT+Feature Disentanglement(FD)+LLM-SER+SED}}: LLaMA SER system fine-tuned to joint SER and SED prediction tasks using disentangled HuBERT representations. 

\vspace{-0.3cm}
\subsection{Evaluation Metrics}
\vspace{-0.2cm}
\noindent{\textbf{SED prediction task}}. To evaluate the accuracy of the SED prediction, we adopt the same supervised metrics in \cite{xu2024secap}, containing standard natural language generation metrics $BLEU@4$($B@4$), $METEOR$($M$), $ROUGE_l$($R$), $CIDEr$($C$), and $SPICE$($S$) \footnote{The $AES_c$ and $AES_s$ metrics in \cite{liang2024aligncap} were produced by GPT-3.5, and not ground truth human labels. They are not used in the paper.}. $B@4$ focuses on the appearance frequency of emotion clues and is used to evaluate the emotional consistency and fine-grainedness of the generated SED. Compared with $B@4$, $M$ considers synonyms more, and $R$ pays more attention to the sufficiency and faithfulness of output. $C$ and $S$ compute the accuracy of SED using human consensus.
\\
\noindent{\textbf{SER and ASR tasks}}. We adopt the commonly used metrics unweighted accuracy (UA) and word error rate (WER) for SER and ASR tasks.

\vspace{-0.3cm}
\subsection{Training Details}
\vspace{-0.2cm}
As mentioned in Sec.\ref{twostages}, we adopt the alternating supervised fine-tuning strategy, with the first and second stages separately conducted for 4 epochs. For the first stage, we set the batch size to 8 and the constant learning rate to $2e^{-4}$ with the first 3\% of steps for warmup. For the second stage, with the same batch size and warmup ratio as the first stage, we adopt a cosine learning rate scheduler and set the peak learning rate to $2e^{-5}$. Additionally, the $\beta$ in the IB regularization iterates from $1e^0$ to $1e^{-4}$, which will be further discussed in Sec. \ref{4.5}.

\begin{table}
    \centering
    \caption{SED prediction performance comparison with baselines and state-of-the-art methods on the EMOSEC dataset. ``P'' stands for the results of methods from public papers. ``$\ddagger$'' represents statistically significant (Paired Single-tailed t TEST \cite{champoux2022first}, $p$=0.05) score improvement over Sys. 1.}
    \label{table1}
    \fontsize{7.4}{10}\selectfont 
    \setlength{\tabcolsep}{1pt} 
    \begin{tabularx}{\columnwidth}{c|c|c|c|c|c|c}
        \Xhline{2\arrayrulewidth}
        \textbf{ID} & \textbf{System} & \textbf{B@4} $\uparrow$ & \textbf{M} $\uparrow$ & \textbf{R} $\uparrow$ & \textbf{C} $\uparrow$ & \textbf{S} $\uparrow$ \\
        \hline
        \multicolumn{7}{c}{\textbf{State-of-the-art SED prediction methods}} \\
        \hline
        P1& HTSAT-BART\cite{mei2024wavcaps} & 4.5 & 11.6 & 20.4 & 5.1 & 3.7 \\
        P2&NoAudioCap\cite{deshmukh2024training} & 6.7 & 14.5 & 21.8 & 10.3 & 5.7 \\
        P3&SECap\cite{xu2024secap} & 7.4 & 16.6 & 25.9 & 11.2 & 5.8 \\
        P4&AlignCap\cite{liang2024aligncap} & 9.8 & 20.9 & 29.8 & 18.7 & 7.6 \\
        \hline
        \multicolumn{7}{c}{\textbf{Baselines \& Proposed method}} \\
        \hline
        1&HuBERT+LLM-SER+SED+ASR & 7.1 & 15.8 & 24.1 & 10.7 & 5.9 \\
        2&HuBERT+FD+LLM-SER+SED & $10.2^{\ddagger}$ & $22.0^{\ddagger}$ & $30.1^{\ddagger}$ & $19.2^{\ddagger}$ & $8.8^{\ddagger}$ \\
        \textbf{3} & \textbf{HuBERT+FD+LLM-SER+SED+ASR} & \textbf{10.3}$^{\ddagger}$ & \textbf{22.1}$^{\ddagger}$ & \textbf{30.4}$^{\ddagger}$ & \textbf{19.3}$^{\ddagger}$ & \textbf{9.0}$^{\ddagger}$ \\
        \Xhline{2\arrayrulewidth}
    \end{tabularx}
\end{table}

\begin{table}[t]
    \caption{SER performance comparison with baselines and state-of-the-art methods on IEMOCAP and MELD datasets. ``--'' means there is no evaluation result on the corresponding task in public papers. ``$\ast$'', ``$\ddagger$'' and ``$\circ$'' represent statistically significant (Paired Single-tailed t TEST \cite{champoux2022first}, $p$=0.05) accuracy improvements over Sys.4, Sys.5 and Sys.6, respectively.}
    \label{table2}
    \setlength{\tabcolsep}{1pt} 
    \fontsize{6.8}{10}\selectfont 
    \centering
        \begin{tabularx}{\columnwidth}{c|c|cc|cc}
            \Xhline{2\arrayrulewidth}
            \multirow{2}{*}{\textbf{ID}} & \multirow{2}{*}{\textbf{System}} & 
            \multicolumn{2}{c|}{\textbf{IEMOCAP}} & \multicolumn{2}{c}{\textbf{MELD}} \\
            \cline{3-6}
            & & \textbf{UA}\% $\uparrow$ & \textbf{WER}\% $\downarrow$ & 
            \textbf{UA}\% $\uparrow$ & \textbf{WER}\% $\downarrow$ \\
            \Xhline{2\arrayrulewidth}
            \multicolumn{6}{c}{\textbf{LLM-based SER methods}} \\
            \hline
            P1 & Whisper+Qformer+LLaMA\cite{tang2023salmonn} & 67.0 & -- & 32.9 & -- \\
            P2 & Whisper+Qwen\cite{chu2023qwen}             & --   & -- & 55.7 & -- \\
            P3 & Whisper+WavLM+LLaMA\cite{hu2024wavllm}    & 72.0 & -- & --   & -- \\
            P4 & Whisper+LLaMA\cite{kyung2024enhancing}    & 77.2 & 14.57 & -- & -- \\
            \hline
            \multicolumn{6}{c}{\textbf{Feature disentanglement-based SER methods}} \\
            \hline
            P5 & Semantic-emotion alignment\cite{wang2024blsp} & 
            \textbf{\uline{78.6}} & -- & 56.4 & -- \\
            P6 & Feature orthogonal constraint\cite{yuan2024}  & 
            74.8 & -- & -- & -- \\
            P7 & HuBERT+Two-stage SFT\cite{mohebbi2024disentangling} & 
            64.8 & 25.6 & -- & -- \\
            P8 & Wav2Vec2+Multi-task learning\cite{hsu2023speech} & 
            77.5 & 22.5 & -- & -- \\
            \hline
            \multicolumn{6}{c}{\textbf{SSL-based SER methods}} \\
            \hline
            1  & HuBERT-large\cite{journals/corr/abs-2106-07447} & 
            64.6 & 23.8 & 53.2 & 45.7 \\
            2  & WavLM-large\cite{chen2022wavlm}                & 
            68.9 & 39.1 & 54.6 & 60.2 \\
            3  & Wav2Vec-large\cite{baevski2020wav2vec}         & 
            69.3 & 28.0 & 54.8 & 41.2 \\
            \hline
            \multicolumn{6}{c}{\textbf{Baselines \& Proposed method}} \\
            \hline
            4  & HuBERT+LLM-SER+SED+ASR                     & 
            73.6 & 24.3 & 56.4 & 29.9 \\
            5  & HuBERT+FD+LLM-SER+ASR                      & 
            70.6 & 18.9 & 53.3 & 29.6 \\
            6  & HuBERT+FD+LLM-SER+SED                      & 
            75.5 & /    & 56.5 & / \\
            \textbf{7}  & \textbf{HuBERT+FD+LLM-SER+SED+ASR}         & 
            77.6$^{\ast}$$^{\ddagger}$$^{\circ}$ & \textbf{\uline{11.1}} & \textbf{\uline{60.1}}$^{\ast}$$^{\ddagger}$$^{\circ}$ & \textbf{\uline{25.3}} \\
            \Xhline{2\arrayrulewidth}
        \end{tabularx}
\end{table}

\vspace{-0.3cm}
\subsection{Main Results}
\vspace{-0.2cm}
\label{4.5}
\noindent{\textbf{SED prediction performance comparison with baselines and state-of-the-art methods}} is shown in Tab. \ref{table1}. The proposed system fine-tuned to joint SER-SED prediction and ASR tasks with feature disentanglement (Sys. 3) consistently outperforms the baseline systems (Sys. 1,2) on the SED prediction performance. Statistically significant increases of $B@4$, $M$, $R$, $C$ and $S$ by up to 3.2\%, 6.3\%, 6.3\%, 8.6\% and 3.1\% absolute (45.1\%, 39.9\%, 26.1\%, 80.4\& and 52.5\% relative) are obtained (Sys. 3 vs. Sys. 1), respectively. Specifically, the highest $B@4$ and $M$ demonstrate feature disentanglement can preserve more SED-related information, enabling the LLM decoder to generate more fine-grained SED, while the highest $R$, $C$ and $S$ scores reveal the accuracy of the predicted SED. Additionally, the best-performing system (Sys. 3) can achieve higher evaluation scores than the state-of-the-art SED methods (Sys. P1-P4), which also shows the efficacy of our approach.

\noindent{\textbf{SER performance comparison with baselines and state-of-the-art methods}} is shown in Tab. \ref{table2}. The same trend can be found in the SER task along with ASR. The proposed system fine-tuned to joint SER-SED prediction and ASR tasks with feature disentanglement (Sys. 7) consistently outperforms the baselines (Sys. 4-6). The highest statistically significant improvement of unweighted accuracy can reach 4.0\% and 3.7\% absolute (5.4\% and 6.6\%) on IEMOCAP and MELD (Sys.7 vs. Sys.4), respectively. Furthermore, Sys. 7 obtains superior unweighted accuracy compared to most published SER methods (Sys. P1-P8) across both IEMOCAP and MELD datasets, with only a marginal performance gap relative to Sys. P5 on IEMOCAP. This demonstrates the strong competitiveness of our approach in SER and ASR tasks.

Tab. \ref{table3} illustrates that removing any of these techniques, or modifying the parameter settings all lead to performance degradation}}, where Sys. 4 is the best performing system in this paper (also as Sys. 3 in Tab. \ref{table1} and Sys. 7 in Tab. \ref{table2}). We can see that: \textbf{1)} The predicted emotion descriptors offer further explainability for SER with improvements of unweighted accuracy by 7.0\% and 6.8\% absolute (9.9\% and 12.8\% relative) on IEMOCAP and MELD (Sys. 4 vs. Sys. 3), respectively. \textbf{2)} Speech content information also contributes to some extent to SER (Sys. 4 vs. Sys. 1), although it is not as effective as emotion descriptors. \textbf{3)} The value of $\beta$ has an effect on SED prediction and SER performance (Sys.4-8), since $\beta$ controls the trade-off between removing redundant information (high $\beta$) and preserving task-related information (low $\beta$) from the HuBERT SSL representations. When $\beta$ falls within the range of $1e^{-3}$ to $e^{-1}$ (Sys.4,6,7), the SER performance remains relatively stable. \textbf{4)} When $\beta$ set to 1 (Sys. 5), IB regularization degrades into a standard VAE. Compared to other systems with IB (Sys. 4,6-8), there is a significant decrease in the performance of Sys. 5, proving IB is effective for this task. \textbf{5)} The alternating SFT method can efficiently mitigate overfitting in ASR tasks to improve SED prediction and SER performance (Sys. 4 vs. Sys. 2).

\begin{table}
    \centering
    \caption{Ablation studies on the importance of $\beta$ in the IB regularization, training strategies and target outputs for SER. ``Two-stage'' denotes training two stages sequentially without alternating. ``$\ast$'' represents statistically significant (Paired Single-tailed t TEST \cite{champoux2022first}, $p$=0.05) accuracy improvement over Sys.4.}
    \label{table3}
    \setlength{\tabcolsep}{0.7pt} 
    \fontsize{5.6}{10}\selectfont 
    \begin{tabularx}{\columnwidth}{c|Y|c|c|c|c|c|c|c|c|c|c|c|c|c}
        \Xhline{2\arrayrulewidth}
        \multirow{3}{*}{\textbf{ID}} & \multicolumn{5}{c|}{\textbf{Experimental Setup}} & \multicolumn{5}{c|}{\textbf{SED Prediction Accuracy}} & \multicolumn{4}{c}{\textbf{SER \& ASR Accuracy}} \\
        \cline{2-15}
        & \multirow{2}{*}{$\mathbf{\beta}$} & \textbf{Training} & \multicolumn{3}{c|}{\textbf{Target Output}} & \multicolumn{5}{c|}{\textbf{EMOSEC}} & \multicolumn{2}{c|}{\textbf{IEMOCAP}} & \multicolumn{2}{c}{\textbf{MELD}} \\
        \cline{4-15}
        & & \textbf{Strategy} & \textbf{Trans.} & \textbf{Des.} & \textbf{Emo.} & \textbf{B@4} $\uparrow$ & \textbf{M} $\uparrow$ & \textbf{R} $\uparrow$ & \textbf{C} $\uparrow$ & \textbf{S} $\uparrow$ & \textbf{UA} $\uparrow$ & \textbf{WER} $\downarrow$ & \textbf{UA} $\uparrow$ & \textbf{WER} $\downarrow$ \\
        \hline
        1 & \multirow{4}{*}{$e^{-2}$} & One-stage & $\usym{2717}$ & \multirow{2}{*}{$\usym{2713}$} & \multirow{8}{*}{$\usym{2713}$} & 10.2 & 22.0 & 30.1 & 19.2 & 8.8 & 75.5$^{\ast}$ & / & 56.5$^{\ast}$ & / \\
        \cline{7-15} \cline{3-3} \cline{4-4}
        2 & & Two-stage & \multirow{7}{*}{$\usym{2713}$} & & & 8.0 & 17.9 & 26.7 & 16.3 & 5.1 & 76.8 & 10.7 & 59.1 & 25.0 \\
        \cline{3-3} \cline{7-15} \cline{5-5}
        3 & & \multirow{6}{*}{Alternating} & & $\usym{2717}$ & & \multicolumn{5}{c|}{/} & 70.6$^{\ast}$ & 18.9 & 53.3$^{\ast}$ & 29.6 \\
        \cline{7-15} \cline{5-5}
        \textbf{4} & & & & \multirow{5}{*}{$\usym{2713}$} & & \textbf{10.3} & \textbf{22.1} & \textbf{30.4} & \textbf{19.3} & \textbf{9.0} & \textbf{77.6} & \textbf{11.1} & \textbf{60.1} & \textbf{25.3} \\
        \cline{2-2} \cline{7-15}
        5 & $e^{0}$ & & & & & 7.8& 16.9& 25.2& 12.4& 6.7& 72.8$^{\ast}$& 17.7& 53.5$^{\ast}$& 28.6 \\
        \cline{2-2} \cline{7-15}
        6 & $e^{-1}$ & & & & & 9.8 & 21.5& 29.6&18.9& 8.6& 77.3& 11.8& 59.7& 26.0 \\
        \cline{2-2} \cline{7-15}
        7 & $e^{-3}$ & & & & & 10.4 & 21.9 & 30.3& 19.4& 8.9& 77.4& 11.3& 60.2& 25.1 \\
        \cline{2-2} \cline{7-15}
        8 & $e^{-4}$ & & & & & 10.0 & 21.7 & 29.6 & 18.6 & 8.5 & 77.1 & 11.6 & 59.8 & 25.8 \\
        \Xhline{2\arrayrulewidth}
    \end{tabularx}
\end{table}
\vspace{-0.4cm}
\section{Conclusion}
\vspace{-0.1cm}
This paper presents a novel end-to-end LLM-empowered explainable SER approach. Fine-grained SED features are disentangled from HuBERT SSL representations via alternating LLM fine-tuning to joint SER-SED prediction and ASR tasks. VAE compressed HuBERT features derived via IB are used to adjust feature granularity. Experiments on the benchmark IEMOCAP and MELD datasets suggest our approach consistently outperforms the comparable LLaMA SER baselines constructed using either {\bf a)} alternating tasks fine-tuning alone; or {\bf b)} features disentanglement only. Statistically significant increases of SER unweighted accuracy by up to 4.0\% and 3.7\% absolute (5.4\% and 6.6\% relative) are obtained. Future research will focus on exploring the differences in fine-grained emotion descriptors when different people express the same emotion.

\section{Acknowledgements}
This research is supported by Hong Kong RGC GRF grant No. 14200220, 14200021, 14200324, Innovation Technology Fund grant No. ITS/218/21, the project of China Disabled Persons Federation No. CDPF2023KF00002, Youth Innovation Promotion Association CAS grant No. 2023119, and the project of Guangzhou Civil Affairs Science and Technology Foundation No. 2022MZK02.

\bibliographystyle{IEEEtran}
\bibliography{mybib}

\begin{thebibliography}{10}
\providecommand{\url}[1]{#1}
\csname url@samestyle\endcsname
\providecommand{\newblock}{\relax}
\providecommand{\bibinfo}[2]{#2}
\providecommand{\BIBentrySTDinterwordspacing}{\spaceskip=0pt\relax}
\providecommand{\BIBentryALTinterwordstretchfactor}{4}
\providecommand{\BIBentryALTinterwordspacing}{\spaceskip=\fontdimen2\font plus
\BIBentryALTinterwordstretchfactor\fontdimen3\font minus \fontdimen4\font\relax}
\providecommand{\BIBforeignlanguage}[2]{{%
\expandafter\ifx\csname l@#1\endcsname\relax
\typeout{** WARNING: IEEEtran.bst: No hyphenation pattern has been}%
\typeout{** loaded for the language `#1'. Using the pattern for}%
\typeout{** the default language instead.}%
\else
\language=\csname l@#1\endcsname
\fi
#2}}
\providecommand{\BIBdecl}{\relax}
\BIBdecl

\bibitem{shen2024emotion}
S.~Shen \emph{et~al.}, ``Emotion neural transducer for fine-grained speech emotion recognition,'' in \emph{ICASSP}, 2024.

\bibitem{naini2024generalization}
A.~R. Naini \emph{et~al.}, ``Generalization of self-supervised learning-based representations for cross-domain speech emotion recognition,'' in \emph{ICASSP}, 2024.

\bibitem{Tzeng2025}
J.-T. Tzeng \emph{et~al.}, ``Noise-robust speech emotion recognition using shared self-supervised representations with integrated speech enhancement,'' in \emph{ICASSP}, 2025.

\bibitem{conf/icassp/DangVNW23}
A.~Dang \emph{et~al.}, ``Emix: a data augmentation method for speech emotion recognition,'' in \emph{ICASSP}, 2023.

\bibitem{xu2024secap}
Y.~Xu \emph{et~al.}, ``Secap: Speech emotion captioning with large language model,'' in \emph{AAAI}, 2024.

\bibitem{yuan2024}
Z.~Yuan \emph{et~al.}, ``Disentanglement network: Disentangle the emotional features from acoustic features for speech emotion recognition,'' in \emph{ICASSP}, 2024.

\bibitem{mohebbi2024disentangling}
H.~Mohebbi \emph{et~al.}, ``Disentangling textual and acoustic features of neural speech representations,'' \emph{arXiv preprint}, 2024.

\bibitem{si2021variational}
S.~Si \emph{et~al.}, ``Variational information bottleneck for effective low-resource audio classification,'' \emph{INTERSPEECH}, 2021.

\bibitem{wu2024multi}
Y.~Wu \emph{et~al.}, ``Multi-modal emotion recognition using multiple acoustic features and dual cross-modal transformer,'' in \emph{ICASSP}, 2024.

\bibitem{ma2024leveraging}
Z.~Ma \emph{et~al.}, ``Leveraging speech ptm, text llm, and emotional tts for speech emotion recognition,'' in \emph{ICASSP}, 2024.

\bibitem{wang2021variational}
D.~Wang \emph{et~al.}, ``Variational information bottleneck based regularization for speaker recognition.'' in \emph{INTERSPEECH}, 2021.

\bibitem{schuller2004speech}
B.~Schuller \emph{et~al.}, ``Speech emotion recognition combining acoustic features and linguistic information in a hybrid support vector machine-belief network architecture,'' in \emph{ICASSP}, 2004.

\bibitem{gao2024speech}
Y.~Gao \emph{et~al.}, ``Speech emotion recognition with multi-level acoustic and semantic information extraction and interaction,'' in \emph{INTERSPEECH}, 2024.

\bibitem{hsu2023}
J.-H. Hsu \emph{et~al.}, ``Speech emotion recognition using decomposed speech via multi-task learning,'' in \emph{INTERSPEECH}, 2023.

\bibitem{makiuchi2021multimodal}
M.~R. Makiuchi \emph{et~al.}, ``Multimodal emotion recognition with high-level speech and text features,'' in \emph{ASRU}, 2021.

\bibitem{xi2022frontend}
Y.-X. Xi \emph{et~al.}, ``Frontend attributes disentanglement for speech emotion recognition,'' in \emph{ICASSP}, 2022.

\bibitem{jin2015speech}
Q.~Jin \emph{et~al.}, ``Speech emotion recognition with acoustic and lexical features,'' in \emph{ICASSP}, 2015.

\bibitem{tang2023salmonn}
C.~Tang \emph{et~al.}, ``Salmonn: Towards generic hearing abilities for large language models,'' \emph{arXiv preprint}, 2023.

\bibitem{chu2023qwen}
Y.~Chu \emph{et~al.}, ``Qwen-audio: Advancing universal audio understanding via unified large-scale audio-language models,'' \emph{arXiv preprint}, 2023.

\bibitem{hu2024wavllm}
S.~Hu \emph{et~al.}, ``Wavllm: Towards robust and adaptive speech large language model,'' \emph{EMNLP}, 2024.

\bibitem{kyung2024enhancing}
J.~Kyung \emph{et~al.}, ``Enhancing multimodal emotion recognition through asr error compensation and llm fine-tuning,'' in \emph{INTERSPEECH}, 2024.

\bibitem{yang2024large}
C.-H.~H. Yang \emph{et~al.}, ``Large language model based generative error correction: A challenge and baselines for speech recognition, speaker tagging, and emotion recognition,'' in \emph{SLT}, 2024.

\bibitem{mei2024wavcaps}
X.~Mei \emph{et~al.}, ``Wavcaps: A chatgpt-assisted weakly-labelled audio captioning dataset for audio-language multimodal research,'' \emph{TASLP}, 2024.

\bibitem{deshmukh2024training}
S.~Deshmukh \emph{et~al.}, ``Training audio captioning models without audio,'' in \emph{ICASSP}, 2024.

\bibitem{liang2024aligncap}
Z.~Liang \emph{et~al.}, ``Aligncap: Aligning speech emotion captioning to human preferences,'' \emph{arXiv preprint}, 2024.

\bibitem{lin2024clap4emo}
W.-C. Lin \emph{et~al.}, ``Clap4emo: Chatgpt-assisted speech emotion retrieval with natural language supervision,'' in \emph{ICASSP}, 2024.

\bibitem{hu2021lora}
E.~J. Hu \emph{et~al.}, ``Lora: Low-rank adaptation of large language models,'' \emph{arXiv preprint}, 2021.

\bibitem{tishby2000information}
N.~Tishby \emph{et~al.}, ``The information bottleneck method,'' \emph{Allerton}, 2000.

\bibitem{alemi2016deep}
A.~Alemi \emph{et~al.}, ``Deep variational information bottleneck,'' \emph{ICLR}, 2017.

\bibitem{kingma2013auto}
D.~P. Kingma \emph{et~al.}, ``Auto-encoding variational bayes,'' \emph{ICLR}, 2014.

\bibitem{xie2021bayesian}
X.~Xie \emph{et~al.}, ``Bayesian learning for deep neural network adaptation,'' \emph{TASLP}, 2021.

\bibitem{hu2021bayesian}
S.~Hu \emph{et~al.}, ``Bayesian learning of lf-mmi trained time delay neural networks for speech recognition,'' \emph{TASLP}, 2021.

\bibitem{journals/corr/abs-2106-07447}
W.~Hsu \emph{et~al.}, ``Hubert: Self-supervised speech representation learning by masked prediction of hidden units,'' \emph{TASLP}, 2021.

\bibitem{9747490}
H.-J. Chang \emph{et~al.}, ``Distilhubert: Speech representation learning by layer-wise distillation of hidden-unit bert,'' in \emph{ICASSP}, 2022.

\bibitem{fang2024llama}
Q.~Fang \emph{et~al.}, ``Llama-omni: Seamless speech interaction with large language models,'' \emph{arXiv preprint}, 2024.

\bibitem{dubey2024llama}
A.~Dubey \emph{et~al.}, ``The llama 3 herd of models,'' \emph{arXiv preprint}, 2024.

\bibitem{jin2024speechcraft}
Z.~Jin \emph{et~al.}, ``Speechcraft: A fine-grained expressive speech dataset with natural language description,'' in \emph{ACM MM}, 2024.

\bibitem{chen2021gigaspeech}
G.~Chen \emph{et~al.}, ``Gigaspeech: An evolving, multi-domain asr corpus with 10,000 hours of transcribed audio,'' \emph{arXiv preprint}, 2021.

\bibitem{thuseethan2022emosec}
S.~Thuseethan \emph{et~al.}, ``Emosec: Emotion recognition from scene context,'' \emph{NEUROCOMPUTING}, 2022.

\bibitem{Busso2008}
C.~Busso \emph{et~al.}, ``{IEMOCAP}: interactive emotional dyadic motion capture database,'' \emph{LANG RESOUR EVAL}, 2008.

\bibitem{poria2018meld}
S.~Poria \emph{et~al.}, ``Meld: A multimodal multi-party dataset for emotion recognition in conversations,'' \emph{arXiv preprint}, 2018.

\bibitem{ma2024emobox}
Z.~Ma \emph{et~al.}, ``Emobox: Multilingual multi-corpus speech emotion recognition toolkit and benchmark,'' \emph{INTERSPEECH}, 2024.

\bibitem{yuan2024disentanglement}
Z.~Yuan \emph{et~al.}, ``Disentanglement network: Disentangle the emotional features from acoustic features for speech emotion recognition,'' in \emph{ICASSP}, 2024.

\bibitem{champoux2022first}
M.-F. Champoux-Larsson \emph{et~al.}, ``How first-and second-language emotion words influence emotion perception in swedish--english bilinguals,'' \emph{BILING-LANG COGN}, 2022.

\bibitem{wang2024blsp}
C.~Wang \emph{et~al.}, ``Blsp-emo: Towards empathetic large speech-language models,'' \emph{EMNLP}, 2024.

\bibitem{hsu2023speech}
J.-H. Hsu \emph{et~al.}, ``Speech emotion recognition using decomposed speech via multi-task learning,'' in \emph{INTERSPEECH}, 2023.

\bibitem{chen2022wavlm}
S.~Chen \emph{et~al.}, ``Wavlm: Large-scale self-supervised pre-training for full stack speech processing,'' \emph{IEEE J. Sel. Top. Signal Process.}, 2022.

\bibitem{baevski2020wav2vec}
A.~Baevski \emph{et~al.}, ``wav2vec 2.0: A framework for self-supervised learning of speech representations,'' \emph{Adv. Neural Inf. Process. Syst.}, 2020.

\end{thebibliography}

\end{document}